\documentclass[11pt,twoside,english,british,oldversion]{article}
\usepackage[T1]{fontenc}
\usepackage[latin9]{inputenc}
\usepackage{color}
\usepackage{babel}
\usepackage{amsmath}
\usepackage{amssymb}
\usepackage{graphicx}
\usepackage[authoryear]{natbib}
\usepackage[unicode=true,pdfusetitle,
 bookmarks=true,bookmarksnumbered=false,bookmarksopen=false,
 breaklinks=false,pdfborder={0 0 1},backref=false,colorlinks=true]
 {hyperref}
\usepackage{breakurl}

\makeatletter

\def\apjl{ApJ}                
\def\apjs{ApJS}               
\def\aap{A\&A}                

\usepackage{amsbsy} 
\usepackage{xcolor}
\usepackage{asp2010}
\resetcounters

\renewcommand{\maketitle}{} 

 \usepackage{textcomp}
\newcommand{\plustimes}{\rlap{+}{\texttimes}}

\makeatother

\begin{document}

\title{Duplicity: its part in the AGB's downfall}

\author{Robert G. Izzard$^{1}$, Denise Keller$^{1,2}$\affil{$^1$Argelander-Institut f\"ur Astronomie, Auf dem H\"ugel 71, 53121 Bonn, Germany}
\affil{$^2$Max-Planck-Institut f\"ur Radioastronomie, Auf dem H\"ugel 69, 53121 Bonn, Germany}}
\maketitle
\begin{abstract}
Half or more of stars more massive than our Sun are orbited by a companion
star in a binary system. Many binaries have short enough orbits that
the evolution of both stars is greatly altered by an exchange of mass
and angular momentum between the stars. Such mass transfer is highly
likely on the asymptotic giant branch (AGB) because this is when a
star is both very large and has strong wind mass loss. Direct mass
transfer truncates the AGB, and its associated nucleosynthesis, prematurely
compared to in a single star. In wide binaries we can probe nucleosynthesis
in the long-dead AGB primary star by today observing its initially
lower-mass companion. The star we see now may be polluted by ejecta
from the primary either through a wind or Roche-lobe overflow. We
highlight recent quantitative work on nucleosynthesis in (ex-)AGB
mass-transfer systems, such as carbon and barium stars, the link between
binary stars and planetary nebulae, and suggest AGB stars as a possible
source of the enigmatic element, lithium. 
\end{abstract}

\section{Duplicitous AGB stars}

The asymptotic giant branch (AGB) phase of stellar evolution is short
relative to the nuclear-burning lifetime of a star. AGB stars are
scarce but, because they are so bright, they are relatively easy to
see. The chance of an AGB star being found in a binary system is also
rare, especially given that AGB stars are bright, variable objects,
so any companion is just hard to spot. There are some spectacular
local examples, e.g.~symbiotic systems such as Mira, but detecting
a secondary star becomes increasingly difficult at longer distance.
Nevertheless, AGB and ex-AGB stars with companions are vitally important
to our understanding of many basic astrophysical processes, see, .e.g.,
the review by Lagadec~\& \foreignlanguage{english}{Chesneau} (\emph{this
volume}). Binary interactions involving AGB stars are crucial to planetary
nebulae, post-AGB stars, white dwarfs, supernovae (especially type~Ia),
gamma-ray bursts, thermonuclear novae and stellar mergers, e.g.~as
gravitational wave sources.

AGB binaries may be rare, but the chemical and dynamical impact of
an AGB star on its companion is commonly observed \citep{1999Jorissen}.
AGB stars are great chemical factories, making elements from the lightest,
such as lithium, carbon and nitrogen, up to the heaviest such as barium,
bismuth and lead. Material rich in these chemicals can be transferred
to a companion star during the short AGB phase. This material pollutes
the companion which retains the chemical signature for the rest of
its life. In a lot of cases, this is for many billions of years. 

Not all stars that are born in binaries are expected to show this
peculiar chemistry from AGB accretion. Close binaries cannot evolve
up the AGB because they undergo mass transfer before or during shell
hydrogen burning on the first giant branch. Very wide binaries have
stars that are so far separated that they do not interact except through
their mutual gravitational attraction. The intermediate cases are
those which interest us: the binary must be wide enough to allow evolution
up the AGB, but close enough that there is mass transfer (probably)
by wind accretion.

Consider a circular binary with a $2\mathrm{\, M_{\odot}}$ primary
and a $0.5\mathrm{\, M_{\odot}}$ secondary star. The primary mass
is deliberately chosen to maximize carbon production at a metallicity
$Z=0.02$ \citep{2010MNRAS.403.1413K}. The timing and mode of mass
transfer depends on the initial separation of the stars. In very close
systems with initial separation less than about $75\mathrm{\, R_{\odot}}$,
mass transfer through Roche-lobe overflow starts on the first giant
branch (using the models of \citealp{Izzard_et_al_2003b_AGBs,2006A&A...460..565I,2009A&A...508.1359I}
as can be found online at \mbox{\href{http://www.astro.uni-bonn.de/~izzard/cgi-bin/binary4.cgi}{www.astro.uni-bonn.de/$\sim$izzard/cgi-bin/binary4.cgi}}).
These binaries may merge, or eject their common envelope \citep{2012IAUS..283...95I,2013A&ARv..21...59I},
but either way will probably not lead to the formation of an AGB binary
system.

At longer separations, up to about $230\mathrm{\, R_{\odot}}$, mass
transfer begins when the primary is on the AGB but before thermal
pulses start. These systems are not expected to be chemically enriched
in, e.g., carbon or $s$-process elements, although they may show
some nuclear burning signature, e.g.~of CN cycling. Again, it is
not clear whether such systems emerge from the common envelope as
a binary or merge. In wider systems than these the primary AGB star
manages to begin thermal pulses, but the system may then suffer mass
transfer by Roche-lobe overflow, likely common-envelope ejection (the
envelope is barely bound) and hence AGB termination. Only wide systems
with separations longer than about $1900\mathrm{\, R_{\odot}}$ avoid
Roche-lobe overflow.

Wide systems can still transfer mass by stellar wind accretion. Whether
this is by classical Bondi-Hoyle-Littleton wind accretion \citep{1944MNRAS.104..273B},
or an enhanced form of mass transfer (e.g.~``wind-Roche-lobe-overflow'',
\citealp{2007ASPC..372..397M}), is a matter of much current debate.
The efficiencies of mass and and angular momentum transfer depend
critically on which mass transfer mode dominates. This then determines
the chemistry in the companion star as well as the final orbital separation
(or period) and eccentricity of the binary. These are properties that
can be observed in binaries such as barium, CH and carbon-enhanced
metal-poor (CEMP) stars. These systems' AGB primaries have long since
shed their envelopes and become white dwarfs, so what we see now is
the (originally) secondary, i.e.~lower mass, star of the binary.
Such systems provide potentially excellent observational constraints
on our binary-star models.

\section{Carbon stars: CH, CEMP, J and R}

AGB stars are expected to be carbon rich, i.e. have $\mathrm{C}/\mathrm{O}>1$
by particle number at the surface, if they have initial masses in
the range $1.5\mathrm{\, M_{\odot}}$ to about $4\,\mathrm{M}_{\odot}$.
Stars in evolutionary phases prior to the AGB should, according to
canonical single-star evolution theory, not be carbon rich. The only
way to make a carbon-rich dwarf or (pre-thermal pulsing AGB) giant
star is then through binary mass transfer. This mass transfer is usually
assumed to be by wind accretion rather than Roche-lobe overflow because
the latter process is thought to lead to rapid ejection of a common
envelope with little accretion on the secondary star. Assuming this
to be the case, systems with $M_{1}=2\mathrm{\, M_{\odot}}$, $M_{2}=0.5\mathrm{\, M_{\odot}}$,
as described above, transfer up to about $0.05\mathrm{\, M_{\odot}}$
by the Bondi-Hoyle mechanism at an initial separation of $2000\mathrm{\, R_{\odot}}$.
The exact amount transferred is subject to considerable uncertainty
and one may question whether the Bondi-Hoyle formalism -- which requires
a wind that is fast relative to the orbital speed -- is at all valid
in AGB stars \citep{2004NewAR..48..843E}. Nevertheless, models of
this type successfully account for the frequency of barium and CH
stars seen in our Galaxy (typically 1\% of G~and~K giants), and
also the paucity of CH stars at (super-)solar metallicity (\citealp{Binary_Origin_low_L_C_Stars};\textbf{
Boyer}~et~al., \emph{this volume}).

Wind accretion models generally fail to reproduce the frequency of
CEMP stars seen in the Galactic halo at metallicities less than $\left[\mathrm{Fe}/\mathrm{H}\right]=-2$.
While observations suggest that CEMPs are 20\% or more of all halo
stars \citep{2013ApJ...762...27Y}, models predict similar fractions
to CH stars, i.e.\ a few per cent \citep{2009A&A...508.1359I}. Resolution
of the CEMP problem is possible by modification of the initial mass
function \citep{2005ApJ...625..833L} however this poses other problems
given the lack of \emph{nitrogen}-enhanced stars \citep{2012A&A...547A..76P}.
Enhancement of carbon in giant molecular clouds from which the stars
formed is a possibility, but it has recently been re-confirmed by
observations that the majority of CEMP stars, which are of the $s$-process
rich CEMP-\emph{s} variety, are binary systems in agreement with the
binary mass-transfer model \citep{2014MNRAS.441.1217S}.

An alternative possibility is to increase the efficiency of wind accretion
which in turn expands the binary-star parameter space available to
CEMP formation. Recent works by the Nijmegen group of Pols, Abate~et~al.
have attempted to do this both statistically and quantitatively \citep{2013A&A...552A..26A}.
Their binary population models are based on those of \citet{2009A&A...508.1359I}
with updates to include wind-Roche-lobe overflow based on the detailed
hydrodynamical models of \citet{2007ASPC..372..397M}. The increased
efficiency of wind accretion compared to earlier works \citep[e.g.][]{2002MNRAS_329_897H}
almost doubles the predicted CEMP frequency, helping to reduce the
discrepancy with observations.

The abundance patterns in individual CEMP stars can now be used to
pin down the initial parameters of their progenitor binary star systems
(Abate\ et\ al. \emph{Astronomy and Astrophysics, submitted}). The
abundance patterns in observed CEMP stars are compared to the binary-star
models using a $\chi^{2}$ technique. This gives best-fitting initial
parameters such as stellar masses and orbital periods. The preliminary
results of this study also show that processes in the secondary star,
such as thermohaline mixing, may not be as efficient as in previous
models \citep[e.g.][]{2007A&A...464L..57S,2009A&A...508.1359I,2009MNRAS.394.1051S}.
This tells us that we do not well understand what happens to material
when it is accreted in a binary system and that our population synthesis
models are lacking some basic physics (e.g. diffusion, gravitational
settling, radiative levitation?).

Close binaries may explain some of the carbon stars found with metallicity
around solar. The R-type stars have identical surface properties to
normal core-helium-burning stars, except that they are carbon rich,
not $s$-process enhanced and are all single \citep{1984ApJS...55...27D,1997PASP..109..256M}.
This is clearly an indication that they were all once binaries that
have merged! Population synthesis studies, such as \citet{2007A&A...470..661I},
show that it is possible to make a sufficient number of merged stars
with the appropriate properties, i.e.~luminosity, temperature, etc.,
\emph{if} carbon, made in the core either during the merger or at
a subsequent helium flash, is mixed to the surface to form the carbon
star. The question then is whether this really happens.

\citet{2010A&A...522A..80P} model stars in the most likely R-star
formation channel from \citet{2007A&A...470..661I}, i.e.~low-mass
red giants merging with helium white dwarfs. They show that a carbon
star does \emph{not} form after the merger. Unfortunately for them,
the most likely formation channel is that with the lowest-mass merged
stars. \citet{2013MNRAS.tmp..656Z} instead model the highest-mass
merger that is predicted by \citet{2007A&A...470..661I} and find
that a carbon star \emph{is }made when the ashes of repeated helium
flashes are mixed into the convective stellar envelope. Most interestingly,
their single star, after the merger, is inflated to roughly the size
of an AGB star. In this inflated phase it appears as a J-type carbon
star (with $\mathrm{C}/\mathrm{O>1}$ and $^{12}\mathrm{C}/^{13}\mathrm{C}<10$
by number). Material expelled during the merging process probably
formed an oxygen-rich disk around the now single star, just as observed
around J stars \citet{2000ApJ...536..438A}. The J star then radiates
its excess thermal energy, shrinks and settles down to core-helium
burning as a carbon-rich, R-type star. This unified model for R and
J stars demonstrates the power of combining binary stellar evolution,
population synthesis and nucleosynthesis, although of course there
are many unanswered questions and uncertainties that remain.

\section{Stellar ejecta: planetary nebulae and the lithium budget}

Binary systems containing an AGB star at periods short enough to enter
Roche-lobe overflow are likely to have their evolution truncated by
common-envelope ejection. It has long been proposed that this induces
the formation of a planetary nebula and some believe that \emph{all}
planetary nebulae are formed by common-envelope ejection \citep{2006ApJ...650..916M}.
If a thermally-pulsing AGB star has its evolution terminated, its
surface chemistry will show \emph{less} of a typical AGB signature,
so -- at low mass -- this should mean less carbon, less \emph{s}-process
elements, but perhaps still some enhancement depending on how many
thermal pulses, with associated third dredge up, occurred.

To test this scenario, we have performed population synthesis calculations
(Keller et~al.\emph{~in preparation}) in which we tag systems as
planetary nebulae when they form either by the classical (single-star)
wind-loss mechanism, or by common-envelope ejection. We find that
most planetary nebulae form through the wind loss channel and only
about 20\% from common-envelope ejection or merger. The majority of
planetary nebulae are AGB stars, although a few per cent may be first
giant branch stars (see also \citealp{2013MNRAS.435.2048H}). 

\begin{figure}
\begin{centering}
\includegraphics[bb=77bp 179bp 586bp 623bp,width=13.4cm]{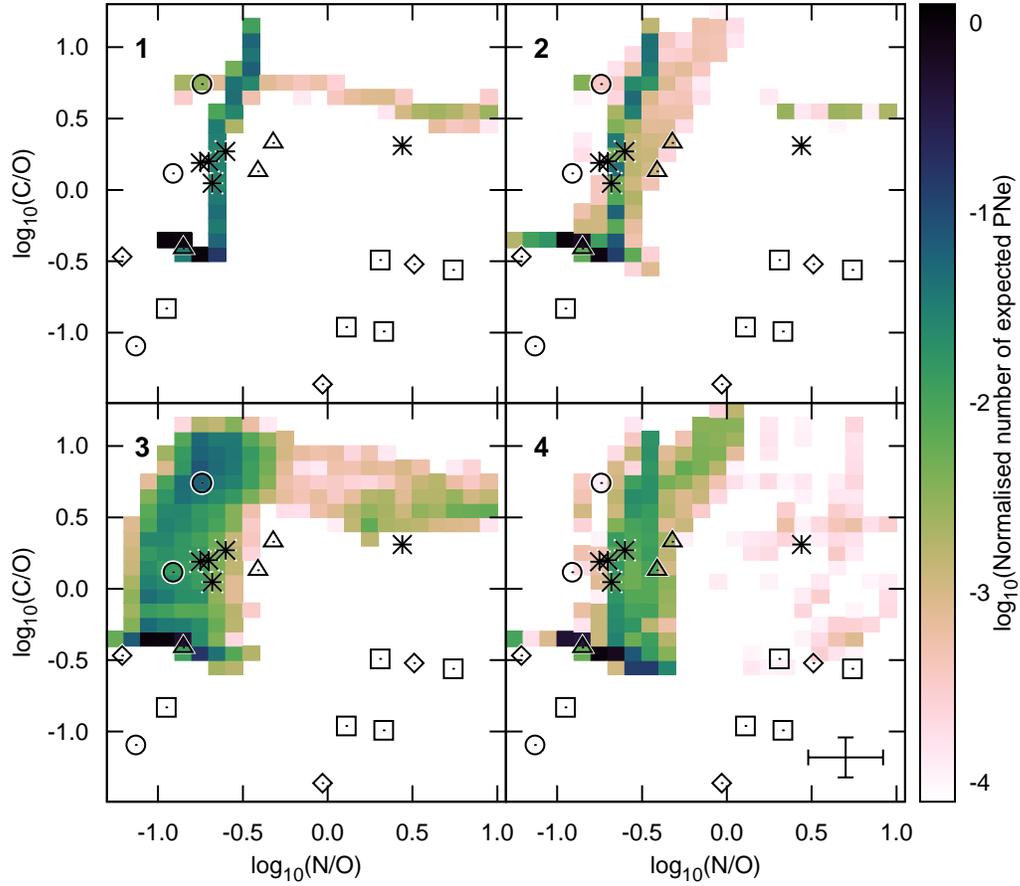}
\par\end{centering}

\protect\caption{\label{fig:Elemental-number-ratios}Elemental number ratios in planetary
nebula progenitors. The panels show four different progenitor channels:
1) single-star wind, 2) binary-star wind (similar to single stars),
3) post-stellar merger wind and 4) common-envelope ejections. Single
stars follow the expected track from detailed AGB models (e.g.\ \citealp{2010MNRAS.403.1413K})
while binaries are smeared out over much more of the parameter space.
The symbols are observed planetary nebulae from \citet{1996A&AS..116...95L,2005ApJ...622..294S}
and \citet{2006A&A...456..451L} which are round ($\odot$), elliptical
(\plustimes), round or elliptical with a bipolar core ($\triangle$),
bipolar ($\square$) and quadrupolar ($\diamondsuit$).}
\end{figure}

Fig.\ \ref{fig:Elemental-number-ratios} shows the elemental ratios
C/O vs N/O (by number of particles) as predicted by our models compared
to observed planetary nebulae. Most planetary nebulae are expected
to have a carbon abundance which is essentially a function of their
AGB progenitor's initial mass, with peak carbon production around
$2\mathrm{\, M_{\odot}}$. Binary companions complicate the picture
not only by truncating the AGB, but also because these stars may merge.
In our model this leads to many nitrogen-rich planetary nebulae because
the extra mass from the merger triggers hot-bottom burning in stars
with mass in excess of $4\mathrm{\, M_{\odot}}$. Caution should be
exercised, however, because we assume that the efficiency of hot-bottom
burning is a function of \emph{total} stellar mass in the merged objects.
This remains to be confirmed by detailed stellar evolutionary models:
the CO core mass may be a more important parameter.

While usually the chemical yield from an AGB star is decreased by
binary-star interaction, there is one element of which the yield may
be increased: lithium. The source of lithium in our Galaxy is unknown
\citep{2012A&A...542A..67P}. AGB stars make lithium if they are massive
enough to process their envelopes by hot-bottom burning. At solar
metallicity this would be stars in the approximate initial-mass range
of $4$ to $8\mathrm{\, M_{\odot}}$ \citep*{2009A&A...508.1343W,2010MNRAS.403.1413K},
where the upper limit is set by the minimum initial mass for a type
II supernova \citep{2014arXiv1410.5431D}. Despite this, AGB stars
are not thought to be a significant source of lithium because once
they make it they very quickly destroy it. Only during the first few
pulses are massive AGB stars rich in lithium, after that they have
very little.

\begin{figure}
\begin{raggedright}
\includegraphics[bb=50bp 50bp 554bp 770bp,angle=270,width=13.4cm]{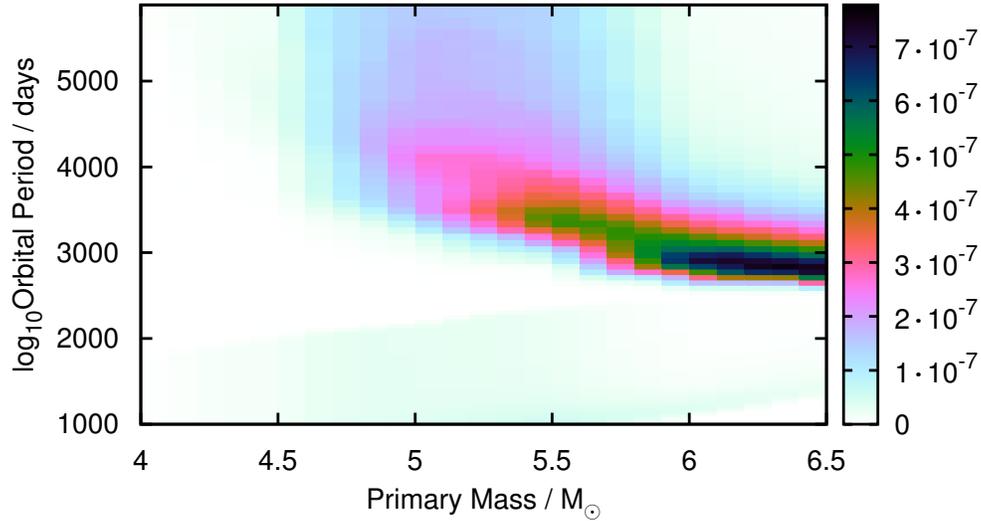}
\par\end{raggedright}

\protect\caption{\label{fig:lithium} Mass ejected as $^{7}\mathrm{Li}$ from a circular
binary-star system with a secondary of initial mass $0.5\mathrm{\, M_{\odot}}$
and metallicity $Z=0.02$ as a function of initial primary mass and
initial orbital period. Equivalent-mass single stars make little lithium
comparable to wide binaries. In binaries in the ``sweet spot'' with
periods around 3000-4000\ days, Roche-lobe overflow starts when the
primary star has made lithium by hot-bottom burning, but not yet destroyed
it. Rapid envelope ejection because of binary-star interaction truncates
the AGB and contributes to the Galactic chemical evolution of lithium. }
\end{figure}

Naturally, this scenario becomes more complicated in binary-star systems.
If mass transfer is initiated at just the right moment, the envelope
-- which is already poorly bound -- should be rapidly ejected in a
common-envelope event. The lithium is then ejected with the envelope,
\emph{not }destroyed, and contributes to the overall Galactic chemical
evolution of lithium. For this scenario to be efficient, the initial
orbital period must be finely tuned such that Roche-lobe overflow
initiates exactly when the stellar lithium abundance is close to maximum.
Fig.\ \ref{fig:lithium} shows this the mass of lithium ejected from
typical binary stars as calculated with the \emph{binary\_c} code
\citep{Izzard_et_al_2003b_AGBs,2006A&A...460..565I,2009A&A...508.1359I}
combined with stellar yields of \citep{2010MNRAS.403.1413K,2014MNRAS.438.1741F}.
Integrating the mass ejected as lithium over all initial masses and
periods, and compared to single stars, binaries make about 25\% more
lithium. While this does not solve the missing lithium problem, it
helps and shows that these binary AGB stars -- while rare -- are important
when considering the chemical evolution of galaxies.

\subsection*{Acknowledgements}

RGI thanks the Alexander von Humboldt foundation for five exhausting
and trying years in Germany. DK is a member of the International Max
Planck Research School (IMPRS) for Astronomy and Astrophysics at the
Universities of Bonn and Cologne and the Bonn-Cologne Graduate School
(BCGS) for Physics and Astronomy and would like to thank the BCGS
for their financial support. 

\bibliographystyle{asp2010}
\bibliography{/export/aibn36_1/izzard/svn/tex/references}

\begin{thebibliography}{}
\expandafter\ifx\csname natexlab\endcsname\relax\def\natexlab#1{#1}\fi
\expandafter\ifx\csname url\endcsname\relax
  \def\url#1{\texttt{#1}}\fi
\expandafter\ifx\csname urlprefix\endcsname\relax\def\urlprefix{URL }\fi
\providecommand{\eprint}[2][]{\url{#2}}

\bibitem[{{Abate} et~al.(2013){Abate}, {Pols}, {Izzard}, {Mohamed}, \& {de
  Mink}}]{2013A&A...552A..26A}
{Abate}, C., {Pols}, O.~R., {Izzard}, R.~G., {Mohamed}, S.~S., \& {de Mink},
  S.~E. 2013, \aap, 552, A26. \eprint{1302.4441}

\bibitem[{{Abia} \& {Isern}(2000)}]{2000ApJ...536..438A}
{Abia}, C., \& {Isern}, J. 2000, \apj, 536, 438.
  \urlprefix\url{http://ukads.nottingham.ac.uk/cgi-bin/nph-bib_query?bibcode=2000ApJ...536..438A&db_key=AST}

\bibitem[{{Bondi} \& {Hoyle}(1944)}]{1944MNRAS.104..273B}
{Bondi}, H., \& {Hoyle}, F. 1944, \mnras, 104, 273

\bibitem[{{Doherty} et~al.(2014){Doherty}, {Gil-Pons}, {Siess}, {Lattanzio}, \&
  {Lau}}]{2014arXiv1410.5431D}
{Doherty}, C.~L., {Gil-Pons}, P., {Siess}, L., {Lattanzio}, J.~C., \& {Lau},
  H.~H.~B. 2014, ArXiv e-prints. \eprint{1410.5431}

\bibitem[{{Dominy}(1984)}]{1984ApJS...55...27D}
{Dominy}, J.~F. 1984, \apjs, 55, 27

\bibitem[{{Edgar}(2004)}]{2004NewAR..48..843E}
{Edgar}, R. 2004, \nar, 48, 843. \eprint{astro-ph/0406166}

\bibitem[{{Fishlock} et~al.(2014){Fishlock}, {Karakas}, \&
  {Stancliffe}}]{2014MNRAS.438.1741F}
{Fishlock}, C.~K., {Karakas}, A.~I., \& {Stancliffe}, R.~J. 2014, \mnras, 438,
  1741. \eprint{1312.1078}

\bibitem[{{Hall} et~al.(2013){Hall}, {Tout}, {Izzard}, \&
  {Keller}}]{2013MNRAS.435.2048H}
{Hall}, P.~D., {Tout}, C.~A., {Izzard}, R.~G., \& {Keller}, D. 2013, \mnras,
  435, 2048. \eprint{1307.8023}

\bibitem[{{Hurley} et~al.(2002){Hurley}, {Tout}, \&
  {Pols}}]{2002MNRAS_329_897H}
{Hurley}, J.~R., {Tout}, C.~A., \& {Pols}, O.~R. 2002, \mnras, 329, 897.
  \urlprefix\url{http://ukads.nottingham.ac.uk/cgi-bin/nph-bib_query?bibcode=2002MNRAS.329..897H&db_key=AST}

\bibitem[{{Ivanova} et~al.(2013){Ivanova}, {Justham}, {Chen}, {De Marco},
  {Fryer}, {Gaburov}, {Ge}, {Glebbeek}, {Han}, {Li}, {Lu}, {Marsh},
  {Podsiadlowski}, {Potter}, {Soker}, {Taam}, {Tauris}, {van den Heuvel}, \&
  {Webbink}}]{2013A&ARv..21...59I}
{Ivanova}, N., {Justham}, S., {Chen}, X., {De Marco}, O., {Fryer}, C.~L.,
  {Gaburov}, E., {Ge}, H., {Glebbeek}, E., {Han}, Z., {Li}, X.-D., {Lu}, G.,
  {Marsh}, T., {Podsiadlowski}, P., {Potter}, A., {Soker}, N., {Taam}, R.,
  {Tauris}, T.~M., {van den Heuvel}, E.~P.~J., \& {Webbink}, R.~F. 2013, \aapr,
  21, 59. \eprint{1209.4302}

\bibitem[{{Izzard} et~al.(2006){Izzard}, {Dray}, {Karakas}, {Lugaro}, \&
  {Tout}}]{2006A&A...460..565I}
{Izzard}, R.~G., {Dray}, L.~M., {Karakas}, A.~I., {Lugaro}, M., \& {Tout},
  C.~A. 2006, \aap, 460, 565

\bibitem[{{Izzard} et~al.(2009){Izzard}, {Glebbeek}, {Stancliffe}, \&
  {Pols}}]{2009A&A...508.1359I}
{Izzard}, R.~G., {Glebbeek}, E., {Stancliffe}, R.~J., \& {Pols}, O.~R. 2009,
  \aap, 508, 1359. \eprint{0910.2158}

\bibitem[{{Izzard} et~al.(2012){Izzard}, {Hall}, {Tauris}, \&
  {Tout}}]{2012IAUS..283...95I}
{Izzard}, R.~G., {Hall}, P.~D., {Tauris}, T.~M., \& {Tout}, C.~A. 2012, in IAU
  Symposium, vol. 283 of IAU Symposium, 95

\bibitem[{{Izzard} et~al.(2007){Izzard}, {Jeffery}, \&
  {Lattanzio}}]{2007A&A...470..661I}
{Izzard}, R.~G., {Jeffery}, C.~S., \& {Lattanzio}, J. 2007, \aap, 470, 661.
  \eprint{arXiv:0705.0894}

\bibitem[{{Izzard} \& {Tout}(2004)}]{Binary_Origin_low_L_C_Stars}
{Izzard}, R.~G., \& {Tout}, C.~A. 2004, \mnras, 350, L1

\bibitem[{{Izzard} et~al.(2004){Izzard}, {Tout}, {Karakas}, \&
  {Pols}}]{Izzard_et_al_2003b_AGBs}
{Izzard}, R.~G., {Tout}, C.~A., {Karakas}, A.~I., \& {Pols}, O.~R. 2004,
  \mnras, 350, 407

\bibitem[{{Jorissen}(1999)}]{1999Jorissen}
{Jorissen}, A. 1999, in IAU Symp. 191: Asymptotic Giant Branch Stars, eds.\ T.
  Le Bertre, A. Lebre, and C. Waelkens, 437

\bibitem[{{Karakas}(2010)}]{2010MNRAS.403.1413K}
{Karakas}, A.~I. 2010, \mnras, 403, 1413. \eprint{0912.2142}

\bibitem[{{Leisy} \& {Dennefeld}(1996)}]{1996A&AS..116...95L}
{Leisy}, P., \& {Dennefeld}, M. 1996, \aaps, 116, 95

\bibitem[{{Leisy} \& {Dennefeld}(2006)}]{2006A&A...456..451L}
--- 2006, \aap, 456, 451. \eprint{astro-ph/0609408}

\bibitem[{{Lucatello} et~al.(2005){Lucatello}, {Gratton}, {Beers}, \&
  {Carretta}}]{2005ApJ...625..833L}
{Lucatello}, S., {Gratton}, R.~G., {Beers}, T.~C., \& {Carretta}, E. 2005,
  \apj, 625, 833. \eprint{arXiv:astro-ph/0412423}

\bibitem[{{McClure}(1997)}]{1997PASP..109..256M}
{McClure}, R.~D. 1997, \pasp, 109, 256

\bibitem[{{Moe} \& {De Marco}(2006)}]{2006ApJ...650..916M}
{Moe}, M., \& {De Marco}, O. 2006, \apj, 650, 916.
  \eprint{arXiv:astro-ph/0606354}

\bibitem[{{Mohamed} \& {Podsiadlowski}(2007)}]{2007ASPC..372..397M}
{Mohamed}, S., \& {Podsiadlowski}, P. 2007, in 15th European Workshop on White
  Dwarfs, edited by {R.~Napiwotzki \& M.~R.~Burleigh}, vol. 372 of Astronomical
  Society of the Pacific Conference Series, 397

\bibitem[{{Piersanti} et~al.(2010){Piersanti}, {Cabez{\'o}n}, {Zamora},
  {Dom{\'{\i}}nguez}, {Garc{\'{\i}}a-Senz}, {Abia}, \&
  {Straniero}}]{2010A&A...522A..80P}
{Piersanti}, L., {Cabez{\'o}n}, R.~M., {Zamora}, O., {Dom{\'{\i}}nguez}, I.,
  {Garc{\'{\i}}a-Senz}, D., {Abia}, C., \& {Straniero}, O. 2010, \aap, 522, A80

\bibitem[{{Pols} et~al.(2012){Pols}, {Izzard}, {Stancliffe}, \&
  {Glebbeek}}]{2012A&A...547A..76P}
{Pols}, O.~R., {Izzard}, R.~G., {Stancliffe}, R.~J., \& {Glebbeek}, E. 2012,
  \aap, 547, A76. \eprint{1209.6082}

\bibitem[{{Prantzos}(2012)}]{2012A&A...542A..67P}
{Prantzos}, N. 2012, \aap, 542, A67. \eprint{1203.5662}

\bibitem[{{Stancliffe}(2009)}]{2009MNRAS.394.1051S}
{Stancliffe}, R.~J. 2009, \mnras, 394, 1051. \eprint{0812.3187}

\bibitem[{{Stancliffe} et~al.(2007){Stancliffe}, {Glebbeek}, {Izzard}, \&
  {Pols}}]{2007A&A...464L..57S}
{Stancliffe}, R.~J., {Glebbeek}, E., {Izzard}, R.~G., \& {Pols}, O.~R. 2007,
  \aap, 464, L57. \eprint{astro-ph/0702138}

\bibitem[{{Stanghellini} et~al.(2005){Stanghellini}, {Shaw}, \&
  {Gilmore}}]{2005ApJ...622..294S}
{Stanghellini}, L., {Shaw}, R.~A., \& {Gilmore}, D. 2005, \apj, 622, 294.
  \eprint{arXiv:astro-ph/0411631}

\bibitem[{{Starkenburg} et~al.(2014){Starkenburg}, {Shetrone}, {McConnachie},
  \& {Venn}}]{2014MNRAS.441.1217S}
{Starkenburg}, E., {Shetrone}, M.~D., {McConnachie}, A.~W., \& {Venn}, K.~A.
  2014, \mnras, 441, 1217. \eprint{1404.0385}

\bibitem[{{Weiss} \& {Ferguson}(2009)}]{2009A&A...508.1343W}
{Weiss}, A., \& {Ferguson}, J.~W. 2009, \aap, 508, 1343. \eprint{0903.2155}

\bibitem[{{Yong} et~al.(2013){Yong}, {Norris}, {Bessell}, {Christlieb},
  {Asplund}, {Beers}, {Barklem}, {Frebel}, \& {Ryan}}]{2013ApJ...762...27Y}
{Yong}, D., {Norris}, J.~E., {Bessell}, M.~S., {Christlieb}, N., {Asplund}, M.,
  {Beers}, T.~C., {Barklem}, P.~S., {Frebel}, A., \& {Ryan}, S.~G. 2013, \apj,
  762, 27. \eprint{1208.3016}

\bibitem[{{Zhang} \& {Jeffery}(2013)}]{2013MNRAS.tmp..656Z}
{Zhang}, X., \& {Jeffery}, C.~S. 2013, \mnras, 656. \eprint{1301.0766}

\end{thebibliography}

\end{document}